\documentclass{iopjournal}
\usepackage{braket}
\usepackage{xcolor}
\usepackage{graphicx}
\usepackage{cite}
\usepackage{babel}
\usepackage{bbm}
\usepackage{amsmath, amsthm, amssymb, mathrsfs}
\usepackage[utf8]{inputenc}

\newcommand{\hilite}{\relax}

\hypersetup{pdfborder = {0 0 0},colorlinks=true,citecolor=magenta,linkcolor=blue,urlcolor=violet}
\begin{document}
\articletype{Paper}

\title{Quantifying Decoherence}

\author{Mohd Shoaib Qureshi\,\orcid{0009-0003-4011-3173} and Tabish Qureshi\,\orcid{0000-0002-8452-1078}}

\affil{Centre for Theoretical Physics, Jamia Millia Islamia, New Delhi, India.}

\email{tqureshi@jmi.ac.in}

\begin{abstract}
Quantum decoherence refers to the phenomenon when the
interaction of a quantum system with its 
environment results in the degradation of quantum coherence. Decoherence
is considered to be the most popular mechanism responsible for the
emergence of classicality from quantum mechanics. The issue of formulating
a measure of decoherence is addressed here. The approach taken here is
that decoherence results from the entanglement of a quantum system with
certain environment degrees of freedom, and quantifying this entanglement
should yield the most natural measure of decoherence. A simple measure of
decoherence is presented based on this notion, and it is examined for
various example systems. The measure proves to be effective and is
relatively straightforward to compute. In addition, a method has been proposed
to measure decoherence in a Mach-Zehnder interferometer \hilite{which may be
useful in neutron interferometry}.
\end{abstract}

\thispagestyle{firstpage}

\section{Introduction}

Emergence of classicality from quantum mechanics is an issue which has been
debated right from its inception.
One approach aimed at tackling this problem employs a form of non-linearity
in quantum evolution, potentially leading to the dynamic evolution of
macroscopic superposition states into a single, distinct macroscopic state
\cite{bassi2013models,carney2021using,universe8020058}. The
origin of this nonlinearity, however, is debatable. One view attributes it
to an inherent non-linearity in the evolution equation
\cite{bassi2003dynamical}. Another considers it resulting from a gravitational
self-interaction \cite{DIOSI1984199,penrose1996gravity,bassi2022}.
The behavior of a quantum system that is weakly interacting with 
numerous of degrees of freedom, referred to as the environment, has been
extensively
examined. Its suggested relationship with the emergence of classicality
has given rise to a vibrant field known as decoherence \cite{zeh,zurek,max}.
The fundamental
concept of the decoherence framework posits that classicality arises as
a property of systems that interact with an environment, which effectively
kills quantum coherence. Both qualitative and quantitative analyses
of decoherence have yielded significant insights into the mechanisms
behind the loss of quantum coherence, and several of its predictions have
been validated through experimental testing \cite{exptdecoh1,brune, monroe}.
Given that decoherence can
disrupt fragile quantum characteristics and, consequently, the operation
of devices that rely on quantum coherence for information processing,
its investigation is of paramount importance for all experimental
applications of quantum information and computation
\cite{monroe,qcompdec1,qcompdec2,qcompdec3}.

Early works on decoherence mostly focused on the time scales over which
decoherence occurs \cite{zurek}, which is interesting for addressing the
question if the emergence of classicality is instantaneous or occurs over
a time scale. With it also arose a need to quantify how classical
a state becomes as it interacts with the environment.
Lee \cite{lee1991} introduced a measure of ``non-classicality"
of a quantum state $\rho$, the quantity $\tau_{\rho}$, referred to as
the ``non-classical depth." This measure was later used by several authors.
The subject of quantifying decoherence has been previously reviewed
\cite{fedichkin,quantrev}. In more recent studies
Zhang and Luo introduced an interesting decoherence measure in terms of
the averaged Wigner-Yanase skew information \cite{luo2021},
Fu and Luo approached quantifying decoherence via increase in classicality
\cite{fu2021}, Gundhi and Ulbricht introduced a quantifier named
``decoherence kernel" \cite{gundhi}.
It has been experimentally demonstrated that partially coherent electron
wave packets can be constructed, which highlights the necessity to develop
methods for quantifying decoherence \cite{metrology}.
\hilite{In an interesting study, decoherence induced during the preparation of
highly non-classical states for quantum metrology, arising from entanglement
with a quantized auxiliary system was studied using quantum Fisher information
(QFI) \cite{qfi}. This approach applies to situations where the auxillary
system cannot be probed, which is also the case with the environment in
standard decoherence. 
However it seems more suited to specific scenarios like state
preparation, and may not apply to, say, environment induced decoherence of
a particle passing through a double slit. Nevertheless, QFI has become a
popular tool to probe decoherence in many subsequent studies.}

\hilite{About a decade back, a specific measure of \emph{coherence} was
established, primarily driven by its necessity in the domain of quantum
information \cite{coherence}. This measure has demonstrated significant
utility, even in quantifying the wave characteristics of quantum particles
\cite{tq15}. One might imagine that, given the strong association between
decoherence and the loss of coherence, one could express decoherence
in terms of the coherence measure. Nevertheless, the challenge in this
approach lies in the fact that the measure of coherence is dependent on
the basis, while it is desirable to have a measure of decoherence
that is independent of the basis, for reasons that will become clear
in the subsequent discussion.}

\hilite{As an open quantum system evolves, it typically establishes
correlations with its environment and irreversibly diminishes its
information content. Nevertheless, this holds true only when the system is
weakly coupled to the environment. If the coupling of the system to the
environment is strong, information may periodically revert to its original
state from the environment, resulting in a back-flow of information. This
phenomenon is referred to as non-Markovian behavior. Of late quantum
non-Markovianity has become an active subject of
research \cite{modi,galve,prabha}, and measures of the degree of
non-Markovianity have been proposed \cite{breuer,guo}.  However, the
decoherence induced emergence of classicality arises from an extremely weak
interaction
with the environment, so minimal that the system effectively evolves
without dissipation. Consequently, non-Markovianity may not be applicable
in these contexts. Therefore, it is necessary to establish a measure that
is relevant to scenarios involving weak interactions with the environment.
}

In this work we consider decoherence as a phenomenon arising out of a quantum
system getting entangled to certain environmental degrees of freedom through
the weakest of interactions. We use this entanglement to quantify the
degree of decoherence of the system.

\section{Decoherence in a nutshell}

The process of decoherence can be generically represented in the following
way. Let the state of a system and its environment be written as
\begin{equation}
|\psi\rangle = \Big(\sum_{j=1}^n c_j |p_j\rangle\Big)|{\cal E}^0\rangle ,
\label{initial}
\end{equation}
where the state of the system is expanded in terms of certain basis states
$|p_j\rangle$, and $|{\cal E}^0\rangle$ is the state of the environment.
Assume a Hamiltonian and time evolution of the form .
\begin{eqnarray}
{\mathscr H} &=& \sum_{j=1}^n |p_j\rangle\langle p_j| \otimes {\mathscr H}^{(j)}\label{decdynamics}\\
{\mathcal U}_t &=& \sum_{j=1}^n |p_j\rangle\langle p_j| \otimes {\mathcal U}_t^{(j)}
\end{eqnarray}
where ${\mathcal U}_t^{(j)} = e^{-i{\mathscr H}^{(j)}t/\hbar}$ and the 
${\mathscr H}^{(j)}$ are certain unspecified Hermitian operators involving
the environment. It is obvious that ${\mathscr H}^{(j)}$ will depend on the
choice of the basis states $|p_j\rangle$.
The state (\ref{initial}), with the above specified time-evolution,
evolves into
\begin{equation}
|\psi(t)\rangle = \sum_{j=1}^n c_j |p_j\rangle {\mathcal U}_t^{(j)}|{\cal E}^0\rangle.
\label{final}
\end{equation}
The state (\ref{final}) represents a grandiose entangled state involving
the system and all of the environment. Practically it is impossible
to keep track of all environmental degrees of freedom, and to probe them
experimentally. That is the reason such a state is rarely used to describe
decoherence. But this entanglement is essentially what leads to decoherence.
This entangled state often leads to a many-worlds type of interpretation
\cite{camilleri}. However, if one wants to look only at the system and
forget about the environment,
it is useful to write the reduced density matrix, which is obtained by
writing the full density matrix for (\ref{final}) as
$\rho_f = |\psi(t)\rangle\langle \psi(t)|$ and tracing over the states
of the environment, and has the form
\begin{equation}
\rho(t) = \sum_{j,k} c_{k}^*c_j |p_j\rangle\langle p_k|
\langle {\cal E}^0|{{\mathcal U}_t^{(k)}}^\dag {\mathcal U}_t^{(j)}|{\cal E}^0\rangle.
\label{reduced}
\end{equation}
Under the kind of time evolution specified above, the diagonal components
of the density matrix, in the basis $|p_j\rangle$, remain unchanged, while
off-diagonal elements are reduced by a factor 
$\langle {\cal E}^0|{{\mathcal U}_t^{(k)}}^\dag
{\mathcal U}_t^{(j)}|{\cal E}^0\rangle \le 1$.
The temporal dependence of the suppressive factors will, in most cases,
rely on the particular model of the environment and its interaction with
the system.
The idea of decoherence is that, for a specific basis $\{|p_j\rangle\}$,
the suppressing terms rapidly decay over short time scales. Thus,
over a time scale, called {\em decoherence time-scale}, the off diagonal
terms in (\ref{reduced}) disappear for all practical purposes, and one
is left with an approximately diagonal density matrix of the system
\begin{equation}
\rho \approx \sum_{j=1}^n |c_j|^2 |p_j\rangle\langle p_j|.
\end{equation}
The off-diagonal elements related to quantum superpositions are now absent.
Consequently, the system \emph{appears to} exhibit classical behavior. This
particular basis is chosen by the interaction with the environment,
resulting in the emergence of classical states. In the literature,
these states are referred to as \emph{pointer states}, and the process of
selection induced by the environment is termed \emph{einselection}.
\vskip 3mm

\section{Quantifying decoherence via entanglement}
\subsection{Entanglement between the system and environment}

From the above discussion it is clear that decoherence is the result of
the system getting entangled with the environment. Our approach is that
the degree of decoherence can be quantified by quantifying the 
entanglement between the system and environment. In this combined 
system-environment aggregate, involving a lot of degrees of freedom,
we consider a bipartition
of the system-environment aggregate such that our system of interest forms
one partition, and the rest of the environment forms the other. One can
now ask, how much is the entanglement between the system and the rest
of the environment. \hilite{To this end we use the methodology developed
by Bhaskara and Panigrahi \cite{bhaskara} for quantifying multipartite
entanglement. Let us assume that our system of interest has interacted
with certain environment degrees of freedom, such that the combined state
of the system and the environment is given by (\ref{final}). We assume
the the degrees of freedom described by the states $|p_i\rangle$ form one
partition, and those described by the states
$\mathcal{U}_t^{(i)}|\mathcal{E}^0\rangle$
form the other much larger partition. We wish to quantify the entanglement
between these two partitions. In the methodology of Bhaskara and Panigrahi
\cite{bhaskara}, one first traces over the degrees of freedom of one partition
to obtain a reduced density operator. In our case, we trace over the
environment states and get the reduced density operator $\rho$ for our system
of interest.} The \emph{generalized concurrence}
for this bipartition can then be written as \cite{bhaskara}
\begin{eqnarray}
E^2 = 4 \sum_{j<k} (\rho_{jj}\rho_{kk} - \rho_{jk}\rho_{kj}) 
 = 2[1 - \text{tr}(\rho^2)] ,
\end{eqnarray}
where $\rho \equiv \sum_i \langle e_i|(|\psi\rangle\langle\psi|)|e_i\rangle
= \text{Tr}_e(\rho_f)$, $\{|e_i\rangle\}$ being an orthonormal basis for
the whole of environment.
This suggests that if the system has a finite dimensional Hilbert space,
we can define a measure of decoherence as
the normalized generalized concurrence for entanglement between the
system and the environment. This decoherence measure is then given by
\begin{equation}
\mathcal{D}_e \equiv \tfrac{n}{n-1}[1 - \text{tr}(\rho^2)] ,
\label{De}
\end{equation}
where $\rho$ is the reduced density operator of the system, after tracing
over the environment states. \hilite{We first test out this measure for the 
scenario described above. In order to get the degree of decoherence of
the system after a time $t$, we plug in the reduced density operator from
(\ref{reduced}) in the expression for the decoherence measure (\ref{De}).
That procedure results in following form of the decoherence measure:}
\begin{eqnarray}
\mathcal{D}_e &=& \tfrac{n}{n-1}[1 - \sum_{i=1}^n \langle p_i|
\sum_{j,k} c_{k}^*c_j |p_j\rangle\langle p_k|
\langle {\cal E}^0|{{\mathcal U}_t^{(k)}}^\dag {\mathcal U}_t^{(j)}|{\cal E}^0\rangle\nonumber\\
&&\sum_{l,m} c_{m}^*c_l |p_l\rangle\langle p_m|
\langle {\cal E}^0|{{\mathcal U}_t^{(m)}}^\dag {\mathcal U}_t^{(l)}|{\cal E}^0\rangle |p_i\rangle\nonumber\\
&=& \tfrac{n}{n-1}\Big[1 - \sum_{i,k} |c_{k}|^2|c_i|^2
|\langle {\cal E}^0|{{\mathcal U}_t^{(k)}}^\dag {\mathcal U}_t^{(i)}|{\cal E}^0\rangle|^2\Big]\nonumber\\
&=& \tfrac{n}{n-1}\Big[1 - \sum_{i=1}^n|c_i|^4 - \sum_{j\neq k} |c_{j}|^2|c_k|^2
|\langle {\cal E}^0|{{\mathcal U}_t^{(k)}}^\dag {\mathcal U}_t^{(j)}|{\cal E}^0\rangle|^2\Big] .
\end{eqnarray}
If the system states do not get correlated to the environment at all, it
implies that \hilite{the environment states correlated to the system states,
$\mathcal{U}_t^{(i)}|\mathcal{E}^0\rangle$, are all identical,
and} $\mathcal{U}_t^{(k)\dag} \mathcal{U}_t^{(j)}=1$ for all
$j,k$. In this situation
$\mathcal{D}_e = \tfrac{n}{n-1}\big[1-\sum_{i=1}^n|c_i|^4 -  \sum_{j\neq k} |c_{j}|^2|c_k|^2\big] = 0$. If the environment states correlated to the states $|p_i\rangle$
are all orthogonal to each other, and $|c_i|^2=\tfrac{1}{n}$ for all $i$,
$\mathcal{D}_e = \tfrac{n}{n-1}[1-\sum_{i=1}^n|c_i|^4] = 1$. This is the
case of maximal decoherence of the system. Here the reduced density operator
of the system is maximally mixed.

\subsection{A qubit coupled to environment of spin-1/2}

In the following we will use the measure introduced in the previous subsection
to a single qubit coupled to a model environment of $N$ spin-1/2 entities.
This system has already been studied in detail by Zurek \cite{zurek82}.
Let the Hamiltonian be represented by
\begin{eqnarray}
H &=& \sum_{j} \epsilon_j |s_j\rangle\langle s_j|
+ \sum_k \mathcal{E}_k|e_k\rangle\langle e_k| \nonumber\\
&& + \sum_{j,k} \gamma_{jk} |s_j\rangle\langle s_j|\otimes|e_k\rangle\langle e_k| ,
\label{Hse}
\end{eqnarray}
where $\epsilon_j$ is the energy eigenvalue of the j'th energy level of
the system, $\mathcal{E}_k$ is the energy of the k'th state of the 
environment, $\gamma_{jk}$ the coupling strength between the respective
states of the system and environment. 
Let the initial system-environment state be:
\begin{eqnarray}
|\Psi(0)\rangle &=& |\psi_0\rangle\otimes|\mathcal{E}_0\rangle\nonumber\\
 &=& \sum_j \alpha_j|s_j\rangle \sum_k \beta_k|e_k\rangle .
\end{eqnarray}
This state will evolve in time via the full Hamiltonian (\ref{Hse}).
The reduced density operator of the system, after a time $t$, is given by
\begin{eqnarray}
\rho(t) = \text{Tr}_e[|\Psi(t)\rangle\langle\Psi(t)|)
 = \sum_{i,j} \rho_{ij}(t)|s_i\rangle\langle s_j| ,
\end{eqnarray}
where
\begin{eqnarray}
\rho_{ii}(t) &=& |\alpha_i|^2\sum_k|\beta_k|^2 = |\alpha_i|^2 \nonumber\\
\rho_{ij}(t) &=& \alpha_i\alpha_j^* e^{-it(\epsilon_i-\epsilon_j)}
\sum_k |\beta_k|^2 e^{-it(\gamma_{ik}-\gamma_{jk})} .
\end{eqnarray}
One can now evaluate the following useful term
\begin{eqnarray}
\text{tr}(\rho^2) 
 &=& \sum_{i,j} |\alpha_i|^2|\alpha_j|^2 \times \nonumber\\
&& \sum_k |\beta_k|^2 e^{-it(\gamma_{ik}-\gamma_{jk})}
 \sum_l |\beta_l|^2 e^{it(\gamma_{il}-\gamma_{jl})} .
\end{eqnarray}
If one define one of the sums in the above expression by
\begin{eqnarray}
 z_{ij} = \sum_k |\beta_k|^2 e^{-it(\gamma_{ik}-\gamma_{jk})} ,
\end{eqnarray}
It has been shown that the time average of it, over a reasonably long
span of time $T$, is \cite{zurek82}
\begin{eqnarray}
 \langle z_{ij}\rangle_T \to 0, \nonumber\\
 \langle|\langle z_{ij}|^2\rangle_T \to \frac{1}{N}. 
\end{eqnarray}
The decoherence measure for the system, at a time $t$, can now be evaluated as
\begin{eqnarray}
\mathcal{D}_e &=& \tfrac{n}{n-1}[1 - \text{tr}(\rho^2)]  \nonumber\\
  &=& \tfrac{n}{n-1}\Big[1 - \sum_i|\alpha_i|^4
- \sum_{i\neq j} |\alpha_i|^2|\alpha_j|^2 \langle|\langle z_{ij}|^2\rangle_T \Big] \nonumber\\
 &\approx& \tfrac{n}{n-1}\Big[1 - \sum_i|\alpha_i|^4 - \frac{1}{N}\sum_{i\neq j} |\alpha_i|^2|\alpha_j|^2\Big] .
\end{eqnarray}
If the environment is large, the last term in the above is negligible,
and the decoherence measure becomes
\begin{eqnarray}
\mathcal{D}_e &\approx& \tfrac{n}{n-1}[1 - \sum_i|\alpha_i|^4] .
\end{eqnarray}
This is the maximum value the decoherence measure will attain, for a given
$\{\alpha_i\}$. If all $|\alpha_i|^2=1/n$, the decoherence measure will attain
its maximal value $\mathcal{D}_e \approx 1$.

\subsection{Decoherence in the spin-boson model}

Next we analyze a model, studied very well in the context of quantum 
dissipation, namely the spin-boson model \cite{sbm}. It involves a
two-level system linearly coupled a to bath
of harmonic oscillators, so that the total Hamiltonian is
\begin{eqnarray}
\mathcal{H} &=& \tfrac{\omega_s}{2}\sigma_z + \sum_k \omega_k b_k^{\dag}b_k 
 + \sum_k \sigma_z(g_k b^{\dag}_k + g_k^* b_k) ,
\end{eqnarray}
where $\sigma$ represents an operator for the two-level system, $b_k$ are
the operators for the harmonic oscillators, and $g_k$ are certain coupling
constants. This model was introduced to study the quantum dissipative
dynamics of a two-level system. However, here we are just concerned with
the decoherence caused by the bath of harmonic oscillators. The effect of
the environment on the two-level system can studied by analyzing the
reduced density operator obtained after tracing over the bath states. The
reduced density operator follows the equation: \cite{sbm}
\begin{eqnarray}
\frac{\partial\rho(t)}{\partial t} &=& -\tfrac{i\omega_s}{2}[\sigma_z,\rho(t)] +
\gamma\sigma_z\rho(t)\sigma_z - \gamma\rho(t) ,
\end{eqnarray}
where $\gamma$ contains all the effect of the heat-bath on the system.
There is a well established procedure to obtain $\gamma$ by assuming a
kind of heat-bath energies and the couplings $g_k$ \cite{sbm}. For an
initial equal superposition of the eigenstates of $\sigma_z$, the reduced
density operator of the system can be obtained as
\begin{eqnarray}
\rho(t) &=& \frac{1}{2}
\begin{pmatrix}
1 & e^{-i\omega_st}e^{-2\gamma t}\\
e^{i\omega_st}e^{-2\gamma t} & 1\\
\end{pmatrix} .
\end{eqnarray}
The decoherence measure can be evaluated from the above to yield
\begin{eqnarray}
\mathcal{D}_e &=& 1 - e^{-4\gamma t} .
\end{eqnarray}
We note that as time progresses, the decoherence measure $\mathcal{D}_e$
approaches its maximal value 1.

\section{Quantifying decoherence for continuous variable systems}

\subsection{Generalized entanglement measure and decoherence}

For continuous variable systems, we follow the same philosophy that we
followed while quantifying decoherence for finite dimensional systems.
We treat the system-environment combine as an entangled system, and divide
into two partitions. The system of interest constitutes one partition, and
the environment constitutes the other. We quantify the entanglement across
this bipartition using the generalized entanglement measure (GEM) introduced
by Swain, Bhaskara, and Panigrahi \cite{swain}.
Consider a $n-$partite entangled state
\begin{eqnarray}
|\psi\rangle = \int \psi(x_1,x_2,\dots,x_n)|x_1\rangle|x_2\rangle\dots|x_n\rangle dx_1 dx_2\dots dx_n ,
\end{eqnarray}
with $\int\psi^*(x_1,x_2\dots,x_n)\psi(x_1,x_2\dots,x_n) d^nx = 1$.
Consider a bipartition $\mathcal{M}$ of the combined Hilbert space, one
containing $m$ degrees of freedom, and the other containing $n-m$ degrees
of freedom. 
The GEM is defined for this $n-$partite entangled state, for the bipartition
$\mathcal{M}$, as \cite{swain}
\begin{eqnarray}
\mathcal{E}_{\mathcal M}^2 &=& 2\Big[1 - \int\Big|\int\psi(x_1',\dots,x_m',y_{m+1},\dots y_n)\nonumber\\
&&\psi^*(x_1,\dots,x_m,y_{m+1},\dots y_n)dy_{m+1}\dots dy_n\Big|^2
d^mx d^mx'\Big],
\label{gemm}
\end{eqnarray}
The GEM is bounded by 0 and 2. For our case of a single system interacting
with a large environment of $n$ degrees of freedom, $m=1$. Since we prefer
a normalized entanglement measure, we drop the 2 in the above expression,
and define the decoherence measure as
\begin{eqnarray}
\mathcal{D}_e &\equiv& \Big[1 - \int\Big|\int\psi(x',y_{1},\dots y_n)
\psi^*(x,y_{1},\dots y_n)dy_{1}\dots dy_n\Big|^2
dx dx'\Big],\nonumber\\
 &=& \Big[1 - \int\big|\rho(x',x)\big|^2 dx dx'\Big],
\label{Dec}
\end{eqnarray}
where $\rho(x',x) = \int\psi(x',y_{1},\dots y_n)
\psi^*(x,y_{1},\dots y_n)d^ny$ is the reduced density matrix
for the system of interest after tracing over the environment states.
In order to evaluate the decoherence measure, we just need to obtain
the reduced density matrix. Eq. (\ref{Dec}) can be written more compactly
as $\mathcal{D}_e = 1 - \text{Tr}[\rho^2]$ which generalizes it to
situations where the system of interest may involve both continuous
and discrete degrees of freedom.

For continuous variable systems, figuring out the pointer basis is a
difficult problem. Various studies have been done on model systems trying
to find out the pointer states \cite{zurek82,zurek1993,av1995,av1998,av1999,paz,wang,gogolin,tq2012}.
The decoherence measure defined by (\ref{Dec}) offers the benefit that
the knowledge of the pointer states is not required for its evaluation.

\subsection{Decoherence of a free particle }

In order to apply the decoherence measure we a consider the simplest
system with continuous variables.  Specifically we consider a particle of mass
$m$, traveling along x-axis. For simply modeling interaction with an
environment, the particle is assumed to be interacting with bath of unit-mass
harmonic oscillators by a coupling of its position to the momenta of the
harmonic oscillators.
If we just look at the motion of the particle in the x-direction, the
Hamiltonian can be written as \cite{savage}
\begin{figure}
\centerline{\resizebox{8.5cm}{!}{\includegraphics{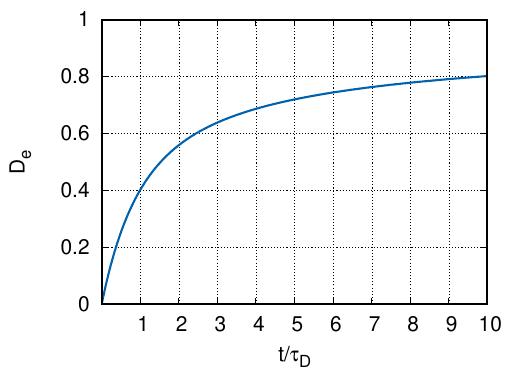}}}
\caption{The decoherence measure $\mathcal{D}_e$ given by (\ref{Decwave}),
plotted against $t/\tau_D$
for a free particle undergoing decoherence, which is initially in a plane
wave state.
}
\label{Dwave}
\end{figure}
\begin{equation}
H = \frac{P_x^2}{2m}+ \sum_k(p_k^2 +
\tfrac{1}{2}\omega_kr_k^2) + x\sum_k g_kp_k ,
\label{Hsw}
\end{equation}
where $x, P_x$ represent the position and momentum operator of the particle,
$r_k, p_k$ are position and momentum of the k'th oscillator. This model
was originally proposed for studying quantum dissipation or damping
\cite{savage}. However, here we use it to study decoherence. The problem
of a particle
coupled to a bath of independent harmonic oscillators was studied in
detail by Caldeira and Leggett \cite{leggett1,leggett2}. One can write a density operator
for the particle and the heat-bath combined, and evolve it using the
Hamiltonian (\ref{Hsw}). The dynamics of the particles can then be studied
by averaging over the environment degrees of freedom, and looking at the
reduced density operator of the particle. Under certain conditions the
reduced density matrix of
the particle follows the equation \cite{leggett1,leggett2}
\begin{eqnarray}
\frac{\partial\rho}{\partial t} &=& \frac{i}{2m\hbar}[P_x^2,\rho]
-\frac{i\gamma}{\hbar}[x,P_x\rho+\rho P_x]
-\frac{4\pi\gamma}{\lambda_T^2}[x,[x,\rho]],
\label{master-sw}
\end{eqnarray}
where $\gamma$ is a damping parameter similar to the one introduced in the
preceding subsection, and $\lambda_T=\sqrt{\frac{2\pi\hbar^2}{mk_BT}}$
is the \emph{thermal deBroglie wavelength} of the particle.
We consider the particle initially to be in a plane wave state 
$\psi(x,0) = e^{ikx}$ with a wave-vector $k$, but confined to a spatially
extended region from $-L$ to $+L$. The initial density matrix of the
particle is $\frac{1}{2L}e^{ik(x-x')}$. With this initial condition the
equation (\ref{master-sw}) can be solved and the reduced density matrix
of the particle at a later time $t$ is given by \cite{savage}
\begin{equation}
\rho(x,x',t) = \exp\{-ie^{-2\gamma t}k(x-x')\}e^{-\sigma(x-x')^2},
\end{equation}
where $\sigma = \frac{\pi}{\lambda_T^2}(1-e^{-4\gamma t})$. Before
proceeding further we would like to emphasize the time scales one should
look at while studying decoherence. In the problem studied by Savage and
Walls \cite{savage}, the parameter $\gamma$ has a simple meaning - it is
the thermal relaxation rate of the particle due to the damping effect of
the environment. However, while studying decoherence one is studying such
minimal effects of the environment that no change in the energy of the
particle results from it, and particle travels virtually undamped
\cite{tq2012,sbrg2007}. The effect of the environment is only a dephasing of
its time evolution. Decoherence effects should come into play at times
much shorter than $1/\gamma$ \cite{tq2012}. For $\gamma t \ll 1$ one can
make the following approximation:
$\sigma \approx \frac{4\pi}{\lambda_T^2}\gamma t$. One can now use
(\ref{Dec}) to evaluate the decoherence measure, and it comes out to be
\begin{equation}
\mathcal{D}_e = 1 - \frac{\sqrt{\pi}}{2\sqrt{2\sigma}L} \text{erf}(\sqrt{2\sigma}L).
\label{Decwave}
\end{equation}
Notice that in the limit $\sqrt{\sigma}L \gg 1$, i.e., for
$t \gg \frac{\lambda_T^2}{4\pi\gamma L^2}$ the decoherence measure
$\mathcal{D}_e$ tends to its maximum value 1. \hilite{Here $1/\sqrt{\sigma}$
is related
to the width of the wavepacket at time $t$. So the limit implies the 
situation where decoherence is strong, and the wavepacket doesn't spread
much even after a time $t$.}
The time scale that emerges
here agrees with the
decoherence time scale $\tau_D=\frac{\lambda_T^2}{4\pi\gamma L^2}$ estimated
earlier \cite{zurek}. For intermediate times one can estimate the degree of
decoherence, the particle has undergone, from (\ref{Decwave}).
The decoherence measure is plotted as a function of time (scaled by $\tau_D$)
in Fig. \ref{Dwave}. It is interesting to see that the growth of the degree
of decoherence with time is distinctly slower than the naive expectation of
$1 - e^{-t/\tau_D}$.

If one starts with a particle which is localized, its wavefunction may be
described by a Gaussian wavepacket 
\begin{equation}
\psi(x,0) = \tfrac{1}{(\pi\sigma^2)^{1/4}} e^{-x^2/2\sigma^2},
\label{psigauss}
\end{equation}
where $\sigma$ is a parameter such that the position spread of the
wavefunction is $2\sigma$. The effect of decoherence on the particle
can be studied by solving (\ref{master-sw}) using the above initial
state. The reduced density matrix of the particle at a time $t$,
in the limit $\gamma t \ll 1$, is then given by \cite{savage}
\begin{eqnarray}
\rho(x,x',t) &=& \tfrac{1}{\sqrt{\pi c}} e^{-ibrR/c},
e^{-R^2/4c} e^{-(a-b^2/c)r^2}
\end{eqnarray}
where $R=x+x',~r=x-x'$, and $a=1/4\sigma^2$, $b=\hbar t/2m\sigma^2$,
$c=\sigma^2+(\frac{\hbar t}{\sigma m})^2 +
\frac{16k_BT\gamma t^3}{m}$. The decoherence measure can now be
evaluated by using (\ref{Dec}), and is  given by
\begin{equation}
\mathcal{D}_e = 1 - \frac{1}{\sqrt{1 + \frac{16k_BT\gamma t^3}{m\sigma^2}}}.
\label{Decgauss1}
\end{equation}
In the limit $t\to 0$, the decoherence measure $\mathcal{D}_e = 0$, 
\hilite{implying that the environment influence is negligible during such short
times}. \hilite{For longer times the environment has enough time to cause
decoherence, so} for
$t \gg (\frac{m\sigma^2}{16k_BT\gamma})^{1/3}$ the decoherence measure 
approaches its maximal value 1. Interestingly, here the decoherence is
not faster if the position spread of the particle is large, i.e., $\sigma$
is large. Rather, it is faster if the momentum spread of the particle
is large, i.e., $\hbar/\sigma$ is large. This is probably an artifact of
the coupling of the particle's position to the momenta of the environment, in
(\ref{Hsw}).
\begin{figure}
\centerline{\resizebox{8.5cm}{!}{\includegraphics{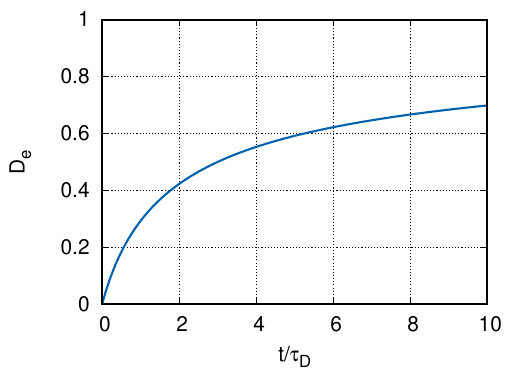}}}
\caption{The decoherence measure $\mathcal{D}_e$ given by (\ref{Decgauss2}),
plotted against $t/\tau_D$
for a free particle undergoing decoherence, which is initially in a Gaussian
state.
}
\label{Dgauss}
\end{figure}

A more realistic coupling to the environment is given by the Hamiltonian
\cite{leggett1,leggett2}
\begin{equation}
H = \frac{P_x^2}{2m}+ \sum_k\left( \tfrac{1}{2}m_k\omega_k^2(r_k -
\tfrac{g_k x}{m_k\omega_k})^2 + \frac{p_k^2}{2m_k}\right) .
\label{Hcl}
\end{equation}
The time evolution of the reduced density matrix is governed by the following
master equation \cite{leggett1,leggett2}
\begin{eqnarray}
\frac{\partial\rho(x,x',t)}{\partial t} &=& \Bigg\{\frac{-\hbar}{2im}
\left(\frac{\partial^2}{\partial x^2} - \frac{\partial^2}{\partial x'^2}\right)
-\gamma(x-x')\nonumber\\
&&\left(\frac{\partial}{\partial x} - \frac{\partial}{\partial x'}\right)
-\tfrac{D}{4\hbar^2}(x-x')^2 \Bigg\} \rho(x,x',t),
\end{eqnarray}
where the parameters are the same as described in the preceding analysis.
Starting from an initial Gaussian state (\ref{psigauss}), the reduced
density matrix of the particle, in the limit $\gamma t \ll 1$, is given
by \cite{tq2012}
\begin{eqnarray}
\rho(R,r,t)
&=&  \tfrac{1}{\sqrt{2\pi\sigma^2}}\exp\left(-\tfrac{2m\gamma k_BTr^2}{\hbar^2}t
 -\tfrac{R^2+r^2}{4\sigma^2}\right) .
\end{eqnarray}
The decoherence measure, evaluated by using (\ref{Dec}), is  given by
\begin{equation}
\mathcal{D}_e = 1 - \frac{1}{\sqrt{1 + t/\tau_D}} ,
\label{Decgauss2}
\end{equation}
where 
\begin{equation}
 \tau_D = \frac{\hbar^2}{2m\gamma k_BT 4\sigma^2} .
\end{equation}
This is consistent with the existing form of the decoherence time-scale,
as the wave-packet is $2\sigma$ wide. However, the increase in decoherence
with time is far from exponential (see Fig. \ref{Dgauss}).

\subsection{Decoherence in a Stern-Gerlach experiment}

Next we look at an experiment which is
a prototype for studying quantum measurement, namely, the Stern-Gerlach
experiment. In this experiment we consider a spin-1/2 particle of mass
$m$, traveling along y-axis, and passing through an inhomogeneous magnetic
field pointing along z-axis. The field is inhomogeneous in the x-direction.
Without going into calculational details, we use this setup to illustrate
the usefulness of the present decoherence measure in dealing with situations
where there are both discrete and continuous degrees of freedom. Two different
kinds of decoherence effects are expected to happen in this experiment, and
at two different time scales. It is interesting to see how the decoherence
measure captures the two effects.
If we just look at the evolution of the particle in the x-direction, the
Hamiltonian can be written as \cite{av1995}
\begin{equation}
H = \frac{p_x^2}{2m}+\lambda\sigma_z +\epsilon x\sigma_z + H_{SE} + H_E ,
\label{Hsg}
\end{equation}
where $x, p_x$ represent the position and momentum operator of the particle,
$\epsilon$ the product of the field gradient and the magnetic moment of the
particle. The environment is assumed to be a bath of independent harmonic
oscillators, with the Hamiltonian $H_E$. The particle interacts with this
environment through an interaction term $H_{SE}$. One writes a density operator
for the particle and the heat-bath combined, and evolves it using the
Hamiltonian (\ref{Hsg}). The dynamics of the particles can then be studied
by averaging over the environment degrees of freedom, and looking at the
reduced density operator of the particle. The reduced density matrix of
the particle follows the equation \cite{av1995}
\begin{eqnarray}
\frac{\partial\rho_{ss'}(x,x',t)}{\partial t} &=& \Bigg\{\frac{-\hbar}{2im}
\left(\frac{\partial^2}{\partial x^2} - \frac{\partial^2}{\partial x'^2}\right)
-\gamma(x-x')\nonumber\\
&&\left(\frac{\partial}{\partial x} - \frac{\partial}{\partial x'}\right)
-\tfrac{\pi\gamma}{\lambda_T^2}(x-x')^2 + \nonumber\\
&&\frac{i\epsilon(xs-x's')}{\hbar} + \frac{i\lambda(s-s')}{\hbar}\Bigg\}
\rho_{ss'}(x,x',t),
\end{eqnarray}
where $s,s'$ are the spin labels $\pm$, $\gamma$ and $\lambda_T$ are the
relaxation rate and the thermal deBroglie wavelength, as discussed in
(\ref{master-sw}).
The effect of the inhomogeneous magnetic field that the particle described
by a localized Gaussian wave-packet evolves into a superposition of two
separated wave-packets, which are correlated the spin states $|\pm\rangle$
defined by $\sigma_z|\pm\rangle=\pm|\pm\rangle$.
In the spin space, the reduced density operator, at a later time $t$, can be
written as
\begin{eqnarray}
\rho(t) &=& 
\begin{pmatrix}
\rho_{++} & \rho_{+-}\\
\rho_{-+} & \rho_{--}\\
\end{pmatrix} .
\end{eqnarray}
It has been argued earlier \cite{zurek} that the decoherence time-scale
$\tau_D$ is inversely proportional to the spatial separation between the
two points between which one is looking at decoherence. In the spin-diagonal
components of the density matrix $\rho_{++}, \rho_{--}$, the separation 
between $x$ and $x'$ is of the order of the width of a Gaussian wave-packet.
On the other hand, in off-diagonal components of the density matrix
$\rho_{+-}, \rho_{-+}$, the separation between $x$ and $x'$ is of the order
of the separation between the two spatially separated Gaussians. Thus
$\rho_{+-}, \rho_{-+}$ decay at a much shorter time scale. Eventually
$\rho_{++}, \rho_{--}$ also experience decoherence, but over a relatively
longer time-scale.

The decoherence measure is given by $\mathcal{D}_e = 1 - \text{Tr}[\rho^2]$,
where the trace now involves a trace over the spin state and a trace over
the spatial degrees of freedom. Partial trace over the spin states of
$\rho^2$ gives
\begin{equation}
 \text{Tr}_{s}[\rho(t)^2] = \rho_{++}^2 + \rho_{--}^2 +
\rho_{+-}\rho_{-+} + \rho_{-+}\rho_{+-} .
\end{equation}
The full trace of $\rho^2$ can then be performed, and the decoherence
measure can be evaluated as
\begin{eqnarray}
{D}_e &=& 1 -\text{Tr}[\rho(t)^2] \nonumber\\
&=& 1 - \int(|\rho_{++}(x,x')|^2
 +|\rho_{--}(x,x')|^2)dxdx' 
- 2\int|\rho_{+-}(x,x')|^2dxdx'
\end{eqnarray}
One can see that over a short time scale the last term in the above will
decay to a negligible value, and the decoherence measure will increase,
but will not reach its maximal value. Over a longer time-scale the
terms involving $\rho_{++}, \rho_{--}$ will also experience a decay, as seen
in the case of decoherence of a Gaussian wavefunction in the preceding
subsection, and the 
decoherence measure will approach its maximal value. The decoherence 
measure is thus well suited to quantify decoherence setting in
progressively over different time scales.

\section{Measuring decoherence}
\begin{figure}
\centerline{\resizebox{8.5cm}{!}{\includegraphics{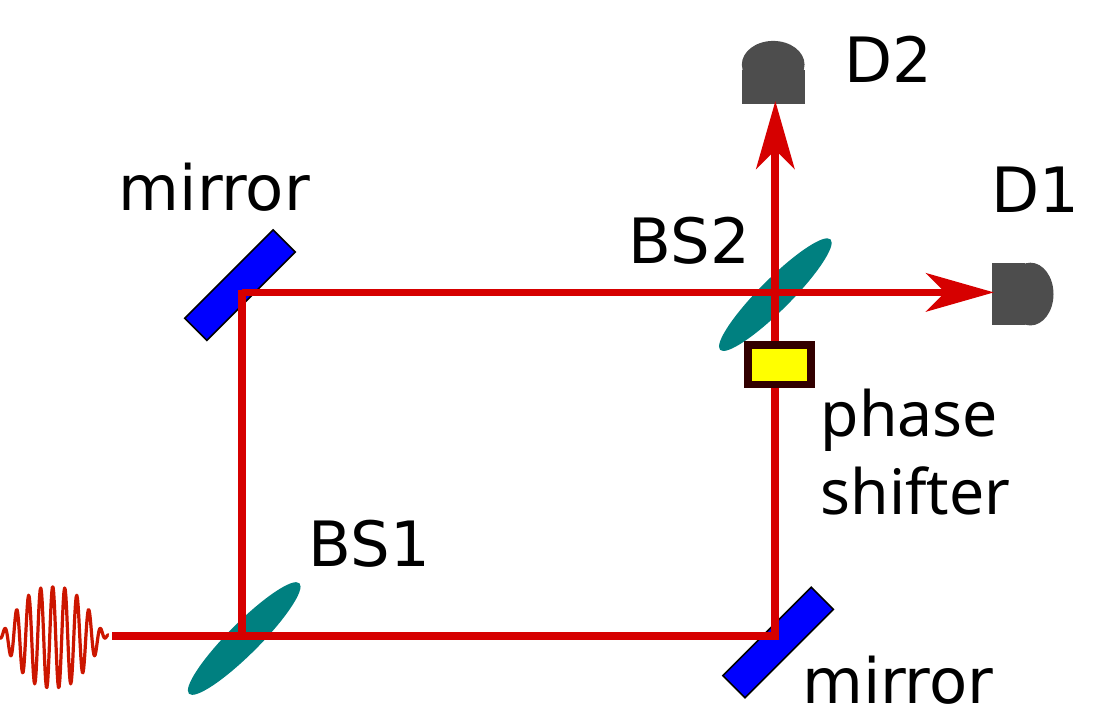}}}
\caption{A schematic diagram of a Mach-Zehnder interferometer. The quanton,
as it traverses the two paths, may undergo decoherence as a result of
interaction with the environment.
}
\label{mz}
\end{figure}

In the following we propose a method of experimentally measuring decoherence
in a Mach-Zehnder interferometer. Consider a Mach-Zehnder interferometer
as shown in Fig. \ref{mz}, where the quantons that travel through it may
be photons, neutrons or any other microparticle. The quanton first encounters
the beam-splitter BS1 and is split into a superposition of two paths.
Let us assume that the state of the quanton, after passing through BS1, is
\begin{equation}
 |\psi_0\rangle = \tfrac{1}{\sqrt{2}}(|\psi_1\rangle + |\psi_2\rangle)
\end{equation}
By the time the quanton reaches the beam-split BS2, the phase difference
between the paths may have changed. We assume an additional phase difference
between the two paths so that the state is now
\begin{equation}
 |\psi\rangle = \tfrac{1}{\sqrt{2}}(|\psi_1\rangle + e^{i\phi}|\psi_2\rangle) .
\label{2paths}
\end{equation}
The effect of BS2 on the quanton is such that it splits a path into a 
superposition of two paths, one traveling to the detector $D1$ and the
other traveling to the detector $D2$. However there is a difference in 
the relative phases of the split paths depending on whether the quanton
came from the upper path (1) or the lower path (2). The action of BS2 can be 
summarized as follows
\begin{eqnarray}
 U_{BS2}|\psi_1\rangle &=& \tfrac{1}{\sqrt{2}}(|D_1\rangle + |D_2\rangle) \nonumber\\
 U_{BS2}|\psi_2\rangle &=& \tfrac{1}{\sqrt{2}}(|D_1\rangle - |D_2\rangle) .
\end{eqnarray}
In optics this happens when the lower surface of BS2 is silvered.
So after passing through BS2, the quanton will be in the state
\begin{eqnarray}
 |\psi_f\rangle &=& U_{BS2}\tfrac{1}{\sqrt{2}}(|\psi_1\rangle + e^{i\phi}|\psi_2\rangle) \nonumber\\
 &=& \tfrac{1}{2}(|D_1\rangle+|D_2\rangle) +
e^{i\phi}\tfrac{1}{2}(|D_1\rangle-|D_2\rangle) .
\end{eqnarray}
The probability of the quanton hitting D1 is $(1+\cos\phi)/2$, and that of
hitting D2 is $(1-\cos\phi)/2$. If $\phi=0$ the probability of the quanton
hitting D2 is zero, and all the quantons will hit D1. This represents sharp
interference. Small nonzero values of $\phi$ lead to unsharp interference.

Now if the quanton, while traversing the two paths, interacts with the
environment, and undergoes decoherence by the time it reaches BS2, it
cannot be represented by a pure state. One can trace over the environment
states, and obtain a reduced density matrix for the quanton. One can look
at the density matrix in the 2-dimensional Hilbert space spanned by 
$|\psi_1\rangle, |\psi_2\rangle$. In this space it looks like
the following:
\begin{eqnarray}
\rho &=& \begin{pmatrix}
\rho_{11} & \rho_{12}e^{-i\phi}\\
\rho_{21}e^{i\phi} & \rho_{22}\\
\end{pmatrix} ,
\label{mixed}
\end{eqnarray}
where $\rho_{11}$ ($\rho_{22}$) represents the probability of the quanton
passing through the upper (lower) path. \hilite{We would like to emphasize
here that although the analysis started with a pure quanton state
(\ref{2paths}), purity of the initial state is not at all a requirement.
The quanton may already be in a mixed state before entering the setup, but
the expression (\ref{mixed}) will still hold. The phase difference $\phi$
does not represent the phase coherence in the initial state. Rather the
phase difference may be introduced in the two paths by the phase shifter
shown in Fig. \ref{mz}.}
The decoherence measure can then
be calculated using (\ref{De}) as
\begin{eqnarray}
{D}_e &=& 2(1 - \rho_{11}^2 -\rho_{22}^2 - 2|\rho_{12}|^2) .
\end{eqnarray}
The problem at hand now is to find a way to make measurements in the experiment
to evaluate the above decoherence measure. If one blocks the lower path
before it enters BS2, there will be equal intensity at D1 and D2 which will
be proportional to $\rho_{11}$. On the other hand, if one blocks the upper path
before it enters BS2, there will again be equal intensity at D1 and D2 which
will now be proportional to $\rho_{22}$.  It can be shown that the average
intensity at the two detectors, in the two scenarios, is given by
\begin{eqnarray}
(I_{av})_1 = \tfrac{1}{2}(I_{D1} + I_{D2})_1 &=& \rho_{11} \nonumber\\
(I_{av})_2 = \tfrac{1}{2}(I_{D1} + I_{D2})_2 &=& \rho_{22} ,
\end{eqnarray}
where $I_{D1},I_{D2}$ represent the intensities at the two detectors, and
the subscripts after the brackets denote which path is open. When both the
paths are open, the intensities at the two detectors are given by
$I_{D1} = \rho_{11} + \rho_{22} + \rho_{12}e^{-i\phi} + \rho_{21}e^{i\phi}$
and $I_{D2} = \rho_{11}+\rho_{22}-\rho_{12}e^{-i\phi}-\rho_{21}e^{i\phi}$.
The difference in the intensities at the two detectors is then given by
\begin{eqnarray}
I_{diff} = \tfrac{1}{2}(I_{D1} - I_{D2}) = 2|\rho_{12}|\cos\phi
\end{eqnarray}
The decoherence measure can now be evaluated as 
\begin{eqnarray}
{D}_e &=& 2[1 - (I_{av})_1^2 - (I_{av})_2^2 - \tfrac{1}{2}I_{diff}^2],
\end{eqnarray}
where we have assumed $\phi=0$. If one is unsure about $\phi$ in the setup,
one can tune the phase shifter such that maximum value of $I_{diff}^2$
is obtained. In this analysis it is assumed that all intensities are normalized
with the total intensity. \hilite{One may mention here that even if the incoming
quanton is not fully coherent, it will not affect results, because each 
quanton is split into a superposition of the two paths by a unitary action
of the beam splitter BS1. Any fluctuation in the phase of the incoming 
quanton will not affect superposition of the two paths, as that initial
phase will factor out of the superposition state. Interference will be
affected only by what happens after BS1. Nevertheless, the results may be
affected by some experimental factors such as detector efficiency, and
other factors depending on the specific setup.}

Given that the Mach-Zehnder interferometer
with cold neutrons is widely used and there have been significant
developments in the field of neutron interferometry \cite{hasegawa2006},
the experiment suggested here should be feasible with neutrons, allowing
for the measurement of decoherence. The investigation of decoherence
through neutron interferometry has been previously suggested \cite{pascazio}.
While decoherence in a conventional
photonic Mach-Zehnder interferometer in a lab would be negligible, one
may use the proposed method to measure decoherence in optical interferometry
involving large distances, e.g., as in the setup involving gravitational
wave detection.
Interferometry with atoms and molecules is a topic of active research,
and decoherence is an important issue there \cite{neon,c60,tqav,horn}.
One can think of extending
the method proposed here to such matter-wave interferometry experiments.

\section{Discussion and conclusions}

To summarize, we formulated a measure of decoherence based on the
quantification of entanglement of a system with the environment. The
measure is given by
$\mathcal{D}_e = \tfrac{n}{n-1}[1 - \text{tr}(\rho^2)]$ for systems with
finite-dimensional Hilbert space, and 
$\mathcal{D}_e = [1 - \text{tr}(\rho^2)]$ for continuous variable systems,
where $\rho$ represents the reduced density operator of the system after
tracing over the environment degrees of freedom. This decoherence measure
has minimum value 0 and maximum value 1. Figuring out the pointer
states that emerge out of decoherence of a system is often a difficult
issue. In one basis, the off-diagonal elements might
diminish more rapidly over time, whereas in another basis, they may also
diminish, albeit at a slower rate \cite{tq2012}. 
The present formulation does not depend on the pointer states. One just
needs to evaluate a trace which is basis independent.
\hilite{In our opinion, a strong point of the present measure is that it is
easily applicable to a wide variety of situations.
On the other hand, QFI  doesn't measure decoherence directly, rather it
measures how sensitive a quantum state is to changes in a chosen parameter.
Often decrease of QFI may come, not from decoherence, but from some unitary
evolution. Without careful modeling, QFI cannot cleanly separate noise
from coherent dynamics.}

Some earlier works treat the classicality of a state as quantifying the
degree of decoherence. In our view, these two are separate things. A
state may be close to a classical state, e.g., a coherent state for a 
harmonic oscillator, but it may not have arisen via decoherence. The
classicality of such a state may not be a measure of decoherence.

We have also proposed a method of measuring decoherence in a Mach-Zehnder
interferometer. The method is straightforward, and just involves measuring
intensities in a few different settings. We believe the measure of decoherence
introduced here will turn out to be useful in quantifying decoherence
in different situations, especially in the quantum information related
experiments.

\hilite{While the entanglement between a system and its environment
generally cannot
be undone, non-Markovianity can throw up interesting situations, especially
in scenarios where the distinction between decoherence and dissipation is
blurred \cite{pernice}. How this decoherence measure holds up in such
situations, needs to be explored further.}

\funding{
MSQ acknowledges financial support from the University Grants Commission,
India, through the Junior Research Fellowship (NTA reference number: 231610153280).}



\end{document}